\documentstyle[12pt,fleqn]{article}
\begin{document}
\newcommand{\nc}{\newcommand}
\newcommand{\rnc}{\renewcommand}
\nc{\al}{\mbox{$\alpha$}} \nc{\ap}{\approx} \nc{\bc}{\begin{center}}
\nc{\ec}{\end{center}} \nc{\beq}{\begin{equation}}
\nc{\eeq}{\end{equation}} \nc{\beqn}{\begin{eqnarray}}
\nc{\eeqn}{\end{eqnarray}} \nc{\beqne}{\begin{eqnarray*}}
\nc{\eeqne}{\end{eqnarray*}} \nc{\be}{\mbox{$\beta$}}
\nc{\bp}{\bar{\mbox{$\phi$}}}\nc{\da}{\dot{A}} \nc{\dq}{\dot{q}}
\nc{\op}{\dot{p}} \nc{\ddq}{\ddot{q}} \nc{\dr}{\dot{\!\vr}}
\nc{\pp}{\dot{\!\vp}} \nc{\de}{\delta} \nc{\eq}{equation\ }
\nc{\Eq}{Eq.\ } \nc{\eqs}{equations\ } \nc{\Eqs}{Eqs.\ }
\nc{\edp}{partial differential equation\ } \nc{\edps}{partial
differential equations\ } \nc{\f}{function\ } \nc{\fs}{functions\ }
\nc{\fa}{\frac{1}{2}} \nc{\fr}{\frac{d}{dt}}
\nc{\fx}[1]{\frac{d}{d\:#1}} \nc{\fh}{Hamiltonian function\ }
\nc{\fl}{Lagrangian function\ } \nc{\fle}{\hs{.5cm}\Longrightarrow
\hs{.5cm}} \nc{\ga}{\mbox{$\gamma$}} \nc{\Ga}{\mbox{$\Gamma$}}
\nc{\h}{Hamiltonian\ } \nc{\hz}{Hamiltonization\ } \nc{\hf}{\hfill}
\nc{\hp}{\hs\parindent} \nc{\hs}{\hspace} \nc{\e}{\left}
\nc{\r}{\right} \nc{\me}{mechanics\ } \nc{\mv}{movement\ }
\nc{\na}{\vec{\nabla}} \nc{\noi}{\noindent}
\nc{\p}{\mbox{$\phi$}}\nc{\paf}[3]{\frac{\pr#1}{\pr#2_#3}}
\nc{\pa}[3]{{\pr#1}/{\pr#2_#3}} \nc{\pac}[3]{\frac{\pr#1}{\pr#2^#3}}
\nc{\pu}[4]{\frac{\pr#1^#2}{\pr#3^#4}}
\nc{\pub}[4]{\frac{\pr#1_#2}{\pr#3_#4}}
\nc{\puf}[4]{{\pr#1_#2}/{\pr#3_#4}}
\nc{\po}[4]{\frac{\pr#1_#2}{\pr#3^#4}}\nc{\pr}[1]{\partial#1}
\nc{\pe}[2]{\frac{\pr#1}{\pr#2}} \nc{\pro}{procedure\ }
\nc{\ql}{\linebreak} \nc{\qf}{\pagebreak} \nc{\Th}{\mbox{$\Theta$}}
\nc{\vr}{\vec{\:r}} \nc{\vp}{\vec{\:p}} \nc{\ra}{\vec{A}}
\nc{\vs}{\vspace} \nc{\va}[3]{{\de#1}/{\de#2_#3}}
\nc{\vac}[3]{\frac{\de#1}{\de#2^#3}}
\nc{\vaf}[3]{\frac{\de#1}{\de#2_#3}}
\nc{\vef}[2]{\frac{\de#1}{\de#2}} \nc{\ve}[2]{{\de#1}/{\de#2}}

\begin{titlepage}
\bc {\LARGE{\bf Direct Hamiltonization}} \vs{1cm}

{\LARGE{\bf  for Nambu Systems}} \vs{2.5cm}

\hs{.5cm} {\bf Maria Lewtchuk Espindola}
\vs{.5cm}

\hs{.5cm} Dpto. de Matem\'atica - CCEN

\hs{.5cm} UFPB \ - \ Cidade Universit\'aria

\hs{.5cm} 58051-970 - Jo\~ao Pessoa - PB

\hs{.5cm} Brazil
\vs{1cm}

\hs{.5cm} 12 May 2008 \ec \vs{2cm}

\noi PACS 03.20/+i
\vs{3cm}

\begin{abstract}

The direct hamiltonization procedure applied to  Nambu mechanical
systems proves that the Nambu mechanics is an usual mechanics
described by only one Hamiltonian. Thus a particular case of
Hamiltonian mechanics. It is also proved that any dynamical system
described by the \eq \ $\dr\,=\,\ra(\vr)$ \ is a Nambu system.
\end{abstract}
\end{titlepage}

\section{Introduction}

\hp The direct \hz procedure developed in a previous paper \cite{HD}
do not depend of the number of \eqs of motion  that describes the
mechanical system and so it can be applied to systems with an even
number of motion equations. The Nambu mechanics \cite{{NAM},{EST}}
leads with these systems and so this procedure can be applied to
Nambu systems.

In section 2 it will be set two theorems. The first one  proves that
the Nambu Mechanics is an usual mechanics described by only one
Hamiltonian and so it cannot be a generalization of the Hamiltonian
mechanics. The second one sets that any dynamical system described
by the motion \eq \ $\dr\,=\,\ra(\vr)$ \ is a Nambu system.

Section 3 gives some examples.

\section{Direct Hamiltonization Procedure Applied to Nambu Systems}

\hp The \eqs of motion in Nambu mechanics \cite{NAM} for the
tridimensional problem are

\beq \dr\,=\,\na\:h\:\times\:\na\:g, \eeq \noi where \
$\vr=(x_1,x_2,x_3)$ \ is the position vector and \ $h$ \ e \ $g$ \
constants of motion.

Nambu defined as Hamiltonians of the system the constants of motion\
$h$ \ e \ $g$ \ \cite{NAM}.

The direct \hz \pro applied to Nambu systems results in a
description of this system by only one Hamiltonian, thus resulting
in an usual mechanics. \vs{.4cm}

\noi {\bf Theorem 1}: \vs{.2cm}

The Nambu \me is an usual \me meanwhile singular. \vs{.2cm}
\vs{.3cm}

\noi Proof:
\vs{.2cm}

Let the \h  assume the following shape: \beq
H\,=\,\vp\cdot\e(\na\:h\:\times\:\na\:g\r)+V(\vr) \eeq where \
$\vp=(p_1,p_2,p_3)$ \ is the momentum of the system and \ $V$ \ an
arbitrary \f of space coordinates. Then from the first set of
canonical \eqs of motion (Hamilton equations): \beq
\dr\,=\,\pe{H}{{\vp}}\,=\,\na\:h\:\times\:\na\:g, \eeq where the
following notation is used
\[\frac{\pr}{\pr{\vec{f}}}\,=\,\e(\frac{\pr}{\pr{f_1}}\:,\,\frac{\pr}
{\pr{f_2}}\:,\,\frac{\pr}{\pr{f_3}}\r)\:.\] Then the \Eq (3) \
recovers motion equations given by \Eq (1). \ \ $\bullet$ \vs{.3cm}

To complete the direct \hz \pro it must be defined the momentum \
$\vp$ \ that is  obtained from the second set of Hamilton \eqs as:
\beq
\pp\,=\,-\;\pe{H}{{\vr}}\,=\,\vp\cdot\e[\frac{\pr}{\pr{\vr}}\:\e(\na\:h\:\times\:
\na\:g\r)\r]+\pe{V}{{\vr}}\:. \eeq

Therefore the \fh \ $H$, given by \Eq (2), provides an
\h description to Nambu mechanics. This \fh is a singular one since it is linear in the momentum \ $\vp$.

This theorem is founded in the direct \hz \pro developed in a
previous paper \cite{HD}.

The \h in the direct \hz \pro \cite{HD} is given
by\footnote{Repeated indexes means sum.}
\[H\,=\,A_i(q,t)\:p_i \hs{2cm}(i=1,2,3),\hs{4.3cm}(2')\] where the \fs \ $A_i$ \ must satisfy the Nambu motion
equations.

The \ $A_i$'s \ are obtained directly from \Eq (1) \ and the Hamilton \eqs
\[\dot{\:x_i}\,=\,\paf{H}{p}{i}\,=\,\e(\na\:h\:\times\:\na\:g\r)_i\,=\,A_i,\]
or
\[\dr\,=\,\ra\,=\,\na\:h\:\times\:\na\:g.  \hs{7.3cm}(3')\]
Then the \h for the Nambu systems is obtained from \Eq (3') \ and
\Eq (2')
\[H\,=\,\e(\na\:h\:\times\:\na\:g\r)\cdot\vp.\]

To this \h can be added an arbitrary \f \ $V(\vr)$ \ without
changing the direct \hz procedure. This addition only modifies the
definition of the momentum \ $\vp$ \ as stated in the direct \hz
\pro\cite{HD}. Therefore the \h for the Nambu systems is given by
\Eq (2). \vs{.4cm}

\noi {\bf Theorem 2}: \vs{.2cm}

Any dynamical system described by the \eqs of motion \beq
\dr\,=\,\ra(\vr) \eeq is a Nambu system. \ Or, equivalently, it is
always possible find \fs \ $h$ \ e \ $g$ \ such that \beq
\ra\,=\,\na\:h\:\times\:\na\:g, \eeq where \ $h$ \ and \ $g$ \ are
constants of motion of the system. \vs{.3cm}

\noi Proof:
\vs{.2cm}

If the \fs \ $h$ \ and \ $g$ \ exists, then
\beqn
\ra\cdot\na\:h\,=\,\e(\na\:h\:\times\:\na\:g\r)\cdot\na\;h\,=\,0; \\
\ra\cdot\na\:g\,=\,\e(\na\:h\:\times\:\na\:g\r)\cdot\na\;g\,=\,0.
\eeqn

The \fs \ $h$ \ and \ $g$ \ must be solutions of this system of \edps, or
\beqne
A_1\:\pe{h}{{x_1}}+A_2\:\pe{h}{{x_2}}+A_3\:\pe{h}{{x_3}}\,=\,0; \\
A_1\:\pe{g}{{x_1}}+A_2\:\pe{g}{{x_2}}+A_3\:\pe{g}{{x_3}}\,=\,0,
\eeqne
whose auxiliar system is
\[\frac{d{x_1}}{A_1}\,=\,\frac{d{x_2}}{A_2}\,=\,\frac{d{x_3}}{A_3}\,=\,
\frac{dh}{0}\,=\,\frac{dg}{0}\:,\]
with the intermediary integrals:
\beqne
u_1\,=\,C_1;\\
u_2\,=\,C_2;\\
h\,\,=\,C_3;\\
g\,\,=\,C_4.
\eeqne

Hence the general solution of the system of \edps is:
\[h\,=\,F_1(u_1,u_2);\]
\beq g\,=\,F_2(u_1,u_2), \eeq where \ $u_1$ \ and \ $u_2$ \ are
constants of motion of  the mechanical system described by (5), as
proved below. \ From (5)
\[\frac{d\:u_m}{dt}\,=\,\pub{u}{m}{x}{i}\,\pe{{x_i}}{t}
\,=\,A_i\,\pub{u}{m}{x}{i}\:,\]
with \ $m=1,2$ \ and \ $i=1,2,3$. \ As \ $u_1$ \ and \ $u_2$ \ are intermediary integrals of (7) and (8) then
\[\frac{d\:u_m}{dt}\,=\,0.\]

From the \Eqs (9) and \Eq (6)
\[\ra\,=\,\na\:h\:\times\:\na\:g\,=\,\na\:F_1\:\times\:\na\:F_2. \hs{5cm}\hf (6')\]
As
\[\na\:F_m\,=\,\pub{F}{m}{x}{i}\,\vec{e_i}\,=\pub{F}{m}{u}{n}\,
\pub{u}{n}{x}{i}\,\vec{e_i}\,=\,\pub{F}{m}{u}{n}\,\na\:u_n,\]
then
\beqne
\ra&=&\e(\pub{F}{1}{u}{m}\,\na\:u_m\r)\;\times\;\e(\pub{F}{2}{u}{n}\,\na\:
u_n\r)\,= \\
   &=&\e[\pub{F}{1}{u}{1}\,\pub{F}{2}{u}{2}-\pub{F}{1}{u}{2}\,
   \pub{F}{2}{u}{1}\,\r]\,\e(\na\:u_1\times\na\:u_2\r)
\eeqne
therefore
\beq
\ra\,=\,\e[\:F_1,F_2\r]_{u_{1},u_{2}}\,\,\e(\na\:u_1\times\na\:u_2\r),
\eeq
where
\[\e[\:F_1,F_2\r]_{u_{1},u_{2}}\,=\,\e[\pub{F}{1}{u}{1}\,\pub{F}{2}{u}{2}-
\pub{F}{1}{u}{2}\,\pub{F}{2}{u}{1}\,\r].\]
Since the most common shape of the Nambu system is:
\beqne
F_1\,\equiv\,u_1;\\
F_2\,\equiv\,u_2,
\eeqne
then
\[\e[\:F_1,F_2\r]_{u_{1},u_{2}}\,=\,\e[\:u_1,u_2\r]_{u_{1},u_{2}}\,=\,1,\]
resulting in
\[\ra\,=\,\na\:u_1\times\na\:u_2.\]

As \ $u_1$ \ and \ $u_2$ \ are constants of motion then they can be
identified with \ $h$ \ and \ $g$ in \Eq (6), completing the prove
of the theorem. \ \ $\bullet$ \vs{.3cm}

The generalization to Nambu systems of greater dimension can be done easily.

\section{Direct Hamiltonization Applied to a Classical Nambu System}

\subsection{Example of the First Theorem}

\hp As an example of direct \hz \pro applied to Nambu systems it
will be considered the rigid rotator \cite{NAM}.

The generalized coordinates of the rigid rotator are the components
of the angular momentum \ $\vec{l}$.

The \eqs of motion can be written as:
\[\dot{\!\vec{l}}\,=\,\na\:h\:\times\:\na\:g,\]
where \ $h$ \ is the half of the  norm square of the angular
momentum
\[h\,=\,\fa\:\e(l_x^{\:2}+l_y^{\:2}+l_z^{\:2}\r),\]
and \ $g$ \ the kinetic energy of the system
\[g\,=\,\fa\e[\frac{l_x^{\:2}}{I_x}+\frac{l_y^{\:2}}{I_y}+\frac{l_z^{\:2}}{I_z}\r]
,\]
$I_x$, \ $I_y$ \ e \ $I_z$ \ being the inertial moments. \ Hence
\ $h$ \ e \ $g$ \ are motion constants.

As
\[\na\:h\,=\,(l_x,l_y,l_z)\,=\,\vec{l}\]
and
\[\na\:g\,=\,\e(\frac{l_x}{I_x}\:,\,\frac{l_y}{I_y}\:,\,\frac{l_z}{I_z}\r)\:,\]
then
\[\dot{\!\vec{l}}\,=\,\e(a_{zy}\:l_y\:l_z,\;a_{xz}\:l_x\:l_z,\;a_{yx}\:l_y\:l_x
\r),\]
where
\[a_{mn}\,=\,\frac{1}{I_m}-\frac{1}{I_n}\:,\]
with \ $m,n=x,y,z$.

And the \h for this system can be obtained from \ (2):
\[H\,=\,\vp\cdot\e(\na\:h\:\times\:\na\:g\r),\]
or
\[H\,=\,a_{zy}\:p_x\:l_y\:l_z+a_{xz}\:p_y\:l_x\:l_z+a_{yx}\:p_z\:l_y\:
l_x.\]

To this \h it can be added an arbitrary \f \ $V(\vr)$.

\subsection{Example of the Second Theorem}

\hp Consider a dynamical system described by the \eqs of motion:
\beqne
\dot{x}\,=\,2\:x\:(z^2\;-\;y^2); \\
\dot{y}\,=\,2\:y\:(x^2\;-\;z^2); \\
\dot{z}\,=\,2\:z\:(y^2\;-\;x^2).
\eeqne

As this is a Nambu system let determine the constants of motion\ $h$
\ e \ $g$, which are solutions of the \edps \ (7) \ and \ (8).

From \Eqs (5)
\[A_k\,=\,\dot{x}_k \hs{1cm} \hf (k=1,2,3),\]
then the auxiliar system for \Eqs (9) are
\[\frac{dx}{2x(z^2-y^2)}\,=\,\frac{dy}{2y(x^2-z^2)}\,=\,
\frac{dz}{2z(y^2-x^2)}\,=\,\frac{dh}{0}\,=\,\frac{dg}{0}\:,\]
with the intermediary integrals
\beqne
u_1&=&x^2+y^2+z^2;\\
u_2&=&x\:y\:z. \eeqne therefore it can be chosen \beqne
h&=&u_2\,=\,x\:y\:z;\\
g&=&u_1\,=\,x^2+y^2+z^2, \eeqne which are motion constants, as it
was proved in theorem \ 2, so that the Nambu system is described by
the \eqs of motion (1).

The development of the \eqs of motion of this system leads to \beqne
\dr&=&\na(x\:y\:z)\:\times\:\na(x^2+y^2+z^2)\,=\\
   &=&(y\:z,\;x\:z,\;x\:y)\times(2\:x,\;2\:y,\;2\:z),
\eeqne
or
\[\dr\,=\,\e(2\:x\;(z^2-y^2),\;2\:y\;(x^2-z^2),\;2\:z\;(y^2-z^2)
\r),\] which can be identified with the initial \eqs of motion of
the system. Therefore this is a Nambu system.

\section {Final Remarks}

\hp The above theorems can be extended to Nambu system described by
 \ $N$ \ constant of motion (the Nambu Hamiltonians) that makes use
 of a greater number of Nambu Hamiltonians (or constants of motion)
 since the direct \hz \pro allows to determine the Hamiltonian \f
 for any mechanical system described only by its \eqs of motion.

Furthermore the use of the direct hamiltonization \pro proves that
the Nambu mechanics cannot be considered as a generalization of
Hamiltonian mechanics as it is a particular case of this
mechanics.\vs{.8cm}

\noi {\bf Acknowledgments} \vs{.2cm}

The author is grateful to Dr. Nelson Lima Teixeira for the suggested
ideas and profitable debates and to Oslim Espindola (in memoria) for
stimulating discussions.

\end{document}